\documentclass[runningheads]{llncs}

\usepackage[T1]{fontenc}
\usepackage[american]{babel}
\usepackage{microtype}
\usepackage{hyperref}
\usepackage{graphicx}
\usepackage{orcidlink}
\usepackage[nolist]{acronym}
\usepackage{algorithm}
\usepackage{algpseudocode}
\floatname{algorithm}{Listing}
\usepackage{amsmath}
\usepackage{cleveref}
\usepackage[framemethod=tikz]{mdframed}
\usepackage{subcaption}

\AtBeginDocument{\renewcommand{\doi}[1]{\href{https://doi.org/#1}{\nolinkurl{#1}}}}

\begin{acronym}
    \acro{llm}[LLM]{Large Language Model}
    \acro{hpc}[HPC]{High-Performance Computing}
    \acro{mcp}[MCP]{Model Context Protocol}
    \acro{gpt}[GPT]{ChatGPT-5.5}
    \acro{opus}[Opus]{Claude Opus 4.7}
    \acro{qwen}[Qwen]{Qwen3-Coder-Next}
\end{acronym}

\definecolor{boxcolor}{rgb}{0.122, 0.435, 0.698}
\newmdenv[
    innerlinewidth=0.5pt,
    roundcorner=4pt,
    linecolor=boxcolor,
    innerleftmargin=5pt,
    innerrightmargin=5pt,
    innertopmargin=5pt,
    innerbottommargin=5pt,
    skipabove=4pt,
    skipbelow=4pt]
{promptbox}

\begin{document}

\title{Generated, Parallel, Scalable? A Study of Agentic AI-Generated Julia Code on Supercomputers}

\titlerunning{A Study of Agentic AI-Generated Julia Code on Supercomputers}

\newif\ifanonymous

\ifanonymous
\else
    \author{
        Linus Bantel\,\inst{1}\,\orcidlink{0009-0009-0247-4748} \and
        Anna-Lena Roth\,\inst{2}\,\orcidlink{0000-0002-6463-3486} \and
        Jonas Posner\,\inst{2}\,\orcidlink{0000-0002-6491-1626} \and
        Dirk Pfl\"uger\,\inst{1}\,\orcidlink{0000-0002-4360-0212}
    }
    \authorrunning{Bantel et al.}
    \institute{
        Institute for Parallel and Distributed Systems, University of Stuttgart, Germany\\
        \email{\{linus.bantel, dirk.pflueger\}@ipvs.uni-stuttgart.de}\\
        \and
        Fulda University of Applied Sciences, Germany\\
        \email{\{anna-lena.roth, jonas.posner\}@cs.hs-fulda.de}
    }
    \hypersetup{
        pdftitle={{Generated, Parallel, Scalable? A Study of Agentic AI-Generated Julia Code on Supercomputers}},
        pdfsubject={AMTE26},
        pdfauthor={Linus Bantel, Anna-Lena Roth, Jonas Posner, Dirk Pfl\"uger},
        pdfkeywords={{LLM, Agentic AI, Julia, HPC}}
    }
\fi

\maketitle

\sloppy
\begin{abstract}
    Julia is increasingly used in \ac{hpc} as a single-language alternative to combining high-level scripting with low-level systems languages, but achieving scalable performance still requires expertise in parallel programming.
\acp{llm} are increasingly used for code generation and are advancing rapidly with each new version.
Yet, existing studies focus on single-shot prompting rather than \emph{agentic} settings, in which an \ac{llm} autonomously plans, generates, and refines code through tool use.

Using an OpenCode-based agent extended with a Julia-documentation \ac{mcp} server, we study agentic generation of parallel Julia code, focusing on task-based execution with \texttt{Dagger.jl}.
We evaluate three \acp{llm}, OpenAI GPT-5.5, Anthropic Claude Opus 4.7, and the open-weight Qwen3-Coder-Next, on three problems with distinct parallel structures: $\pi$ approximation, tiled general matrix multiplication, and tiled Cholesky decomposition.
The generated \texttt{Dagger.jl} implementations are compared against agent-generated \texttt{Base.Threads} and \texttt{MPI.jl} baselines, with shared-memory experiments scaling to 192 cores and distributed-memory experiments on two nodes.

The agents reliably produce executable code for small inputs but fail at larger scales due to deadlocks, oversubscription, or out-of-memory errors, with the open-weight model affected most severely.
The two commercial models scale comparably on \texttt{Base.Threads} and \texttt{MPI.jl}, while their \texttt{Dagger.jl} implementations expose recurring weaknesses in task dependencies, granularity, and scheduling.
Agentic AI is promising for producing parallel Julia code, but generating robust, performance-aware implementations for large-scale \ac{hpc} systems remains an open challenge.

\end{abstract}

\section{Introduction}\label{sec:intro}
Scientific software development in \ac{hpc} often relies on multiple programming languages with different levels of abstraction.
While high-level languages such as Python, MATLAB, and R support intuitive problem formulation, performance-critical components are commonly implemented in lower-level languages such as C, C++, or Fortran, frequently combined with parallel programming models such as MPI, OpenMP, or CUDA to efficiently exploit modern \ac{hpc} architectures.
This separation creates the \emph{two-language problem}, in which productive high-level code must be reimplemented for performance, increasing expertise requirements and maintenance effort while reducing code reusability~\cite{eschle2023julia,bezanson2017numerical,bezanson2017dynamism}.

Julia addresses this problem by combining a dynamic, interactive syntax suited to numerical and scientific computing~\cite{stewart2025JuliaHEP,byrne2021mpijl} with LLVM-based JIT compilation to efficient native code, achieving performance close to traditional \ac{hpc} languages for numerical kernels~\cite{lin2021comparing,godoy2023evaluating}.
Recent studies show that Julia is competitive with established parallel programming models such as OpenMP, Kokkos, OpenCL, CUDA, HIP, and SYCL on selected \ac{hpc} benchmarks~\cite{lin2021comparing}, and that \texttt{MPI.jl} matches C MPI for most collective operations~\cite{byrne2021mpijl,hunold2020benchmarking}.

While these results demonstrate Julia's potential for \ac{hpc}, scalable performance on modern systems still requires applications to expose and coordinate parallelism across threads, processes, and distributed compute nodes.
Using established models such as MPI or CUDA forces developers to manage communication, synchronization, data movement, and load balancing explicitly, reintroducing complexity despite Julia's high-level model.
Asynchronous Many-Task (AMT) programming models provide a higher-level abstraction by decomposing computations into fine-grained tasks connected via dependencies, with scheduling delegated to a runtime system.
This is particularly attractive for irregular or dynamic workloads, where parallelism is difficult to express through static loop-based or message-passing approaches.
In the Julia ecosystem, \texttt{Dagger.jl}~\cite{alomairy2024dagger} implements this idea by representing computations as directed acyclic graphs that can be scheduled across local and distributed workers.

In parallel, \acp{llm} have become powerful tools for natural language understanding, reasoning, and code generation, drawing growing attention in the \ac{hpc} community.
Available models include OpenAI's GPT family\cite{openai_gpt55}, Anthropic's Claude\cite{claude}, and open-weight alternatives such as Qwen~\cite{qwen}, with new versions appearing rapidly.
Recent work has evaluated \acp{llm} on \ac{hpc}-specific challenges~\cite{bantel2026promptsPerformance,nichols2024large,diehl2024evaluating,diehl2025llama2}, but mostly in single-shot prompting rather than \emph{agentic} settings, where an \ac{llm} plans, writes, executes, and refines code via tools.

One effort to systematize such evaluations across rapidly evolving models is ParEval~\cite{nichols2024large}, which benchmarks \acp{llm} on parallel code generation. 
Its results show strong performance on boilerplate and common patterns, but weaknesses in concurrency control, synchronization, and performance optimization, indicating that \acp{llm} currently complement rather than replace expert knowledge.

In this work, we study \ac{llm}-generated parallel Julia code in an \emph{agentic} setting, focusing on task-based execution with \texttt{Dagger.jl}.
Building on ParEval's prompt design, we adapt its evaluation to this setting.
We compare two commercial models, OpenAI GPT-5.5\cite{openai_gpt55} and Anthropic Claude Opus 4.7\cite{claude}, with the open-weight Qwen3-Coder-Next (35B active parameters)~\cite{qwen}, covering trade-offs in capability, deployment flexibility, and kernel generation time.
All models are driven by an OpenCode-based agent extended with a Julia-documentation \ac{mcp} server.
We consider three representative algorithmic problems with distinct parallel structures, namely $\pi$ approximation via numerical integration (embarrassingly parallel), tiled general matrix multiplication (regular structured), and tiled Cholesky decomposition (irregular dependencies), and compare agent-generated implementations across \texttt{Base.Threads}, \texttt{MPI.jl}, and \texttt{Dagger.jl}.
All experiments run on the Otus supercomputer, using up to 192 cores on one node for shared memory and up to 384 cores across two nodes for distributed memory.
Our results show that the two commercial models reliably produce executable, reasonably scaling code for the established \texttt{Base.Threads} and \texttt{MPI.jl} baselines, while the open-weight model frequently fails at the problem sizes required for meaningful scaling.
Across all models, \texttt{Dagger.jl} implementations reveal recurring weaknesses in task dependencies, granularity, and scheduling, with several runs failing at higher core counts due to deadlocks or out-of-memory errors.

\Cref{sec:related} presents related work, \Cref{sec:setup} the experimental setup, and \Cref{sec:results} the results.
\Cref{sec:discussion} discusses the findings, and \Cref{sec:conclusion} concludes.

\section{Related Work}\label{sec:related}
Generating parallel code for \ac{hpc} using \acp{llm} is challenging, as correctness alone is insufficient; performance and efficiency are equally critical~\cite{ljaljevic2026survey}.
Domain-specific approaches such as HPC-GPT~\cite{ding2023hpcgpt}, MonoCoder~\cite{kadosh2024monocoder}, LASSI~\cite{dearing2024lassi}, and CodeRosetta~\cite{tehranjamsaz2024coderosetta} investigate integrating \acp{llm} into \ac{hpc} workflows to improve productivity, support domain-specific programming, and enable code translation.
However, their effectiveness remains inconsistent for complex parallel workloads.

Recent work examining \ac{llm} capabilities across parallel programming frameworks such as MPI, OpenMP, CUDA, and Kokkos identifies a pronounced parallelism gap: while models perform well on serial tasks, they frequently struggle with data dependencies, synchronization, and thread safety~\cite{nichols2024large}.
Broader multi-language evaluations further show that \acp{llm} can generate functional code and accompanying unit tests for relatively simple scientific applications, but their performance degrades for parallel and distributed workloads~\cite{diehl2025llama2}.
Similarly, studies on performance optimization find that current models can identify obvious anti-patterns yet often fail to apply architecture-specific optimizations that require deeper hardware awareness~\cite{cui2025llmhpc}.
At the same time, specialized models such as CUDA-LLM~\cite{chen2025cudallm} demonstrate that high efficiency can be achieved when models are tailored to specific architectures, particularly for GPU kernel generation.

Within this broader landscape, task-based parallelization in Julia remains underexplored.
Evaluations indicate that \acp{llm} can generate usable code for mature Julia programming models such as \texttt{Base.Threads} and \texttt{CUDA.jl}, but confirm that parallel code generation is substantially more challenging than serial scientific programming~\cite{diehl2024evaluating,godoy2024llm}.
Julia-native task runtimes are therefore natural targets for \ac{llm}-assisted parallelization, as their APIs expose dependency information explicitly, e.g., through Dagger's \texttt{Datadeps} annotations, while remaining close to the mathematical code users intend to write~\cite{alomairy2024dagger,godoy2024llm}.
To the best of our knowledge, no prior work has systematically studied \ac{llm}-based agents for task-based parallelization in Julia, particularly for performance-oriented refinement.

\section{Experimental Setup}\label{sec:setup}
We describe the algorithmic problems (\Cref{subsec:algorithmic-problems}), the evaluated \acp{llm} and code-generation environment (\Cref{sec:setup_gen}), and our two-stage workflow (\Cref{sec:setup_codegen}).

\subsection{Algorithmic Problems}\label{subsec:algorithmic-problems}
Our initial objective was to reproduce the full ParEval benchmark~\cite{nichols2024large} and adapt its tasks and prompts for parallel, task-based Julia, replacing manual prompting with \ac{llm}-based agents that also test and optimize the generated code themselves.
However, a preliminary evaluation revealed two limitations.
First, ParEval's prompts are primarily designed for completion-based systems, such as GitHub Copilot-style autocompletion, rather than for agentic systems that iteratively plan, generate, and revise code.
Second, applying the full 60-experiment benchmark to agentic systems would require substantial time and token budgets, and extensive manual analysis to assess correctness, scalability, and code quality.
We therefore restrict our evaluation to the following three representative algorithmic problems.

\textbf{Approximation of $\pi$ via numerical integration} is an embarrassingly parallel problem.
The integration domain can be divided into independent subintervals, where each task computes a partial result that is aggregated to obtain the final approximation.
Its computational complexity is $\mathcal{O}(1/d)$, where $d$ is the step width of the numerical integration.
Since no data movement is required beyond the final aggregation, the problem is strictly compute-bound.
The problem description provided to the agent is:
\begin{promptbox}
    Approximate pi using the numerical integration $f(x) = 4 / (1 + x^2)$ from $0$ to $1$: $\int_0^1 4/(1+x^2) dx = 4 * \arctan(1) = 4 * \pi/4 = \pi$.
    The function returns the approximated value.
\end{promptbox}

\textbf{Matrix-Matrix Multiplication (GEMM)} is a regular, structured workload with predictable communication patterns and substantial arithmetic intensity per task.
Its tiled formulation maps naturally to task-based execution, and the use of optimized BLAS kernels is explicitly permitted for the agents.
\begin{promptbox}
    Multiply the matrix A by the matrix B.
    Store the results in the matrix C. A is an MxK matrix, B is a KxN matrix, and C is an MxN matrix.
    The matrices are stored in column-major order.
\end{promptbox}

\textbf{Cholesky decomposition} is a standard benchmark for task-based runtime systems.
In contrast to embarrassingly parallel workloads, it exhibits structured task dependencies and therefore serves as a fitting workload for task-based scheduling frameworks.
Typical parallel implementations follow a block-based decomposition, in which the factorization is divided into smaller matrix blocks that can be processed concurrently while respecting data dependencies.
We therefore expect task-based frameworks such as \texttt{Dagger.jl} to be well suited to this workload, as their scheduling mechanisms can exploit task parallelism while managing inter-task dependencies.
\begin{promptbox}
    Factorize the symmetric positive definite matrix A into A = L * L$^T$ where L is a lower triangular matrix.
    The lower triangular part (including the diagonal) contains L.
    The upper triangular part can be left unchanged or ignored.
    A is an NxN matrix stored in column-major order.
    Use POTRF, TRSM, SYRK, and GEMM block-based.
\end{promptbox}

\subsection{\acp{llm} and Code Generation Environment}\label{sec:setup_gen}
We evaluate three \acp{llm}, each assessed independently under identical conditions for a controlled comparison.

\textbf{\ac{gpt}}~\cite{openai_gpt55} is a proprietary general-purpose \ac{llm} provided by OpenAI.
Since the model is commercial, local deployment and direct inspection of weights or architectural details are not possible.
We access it through OpenAI's ChatGPT Plus subscription plan.

\textbf{\ac{opus}}~\cite{claude} is a proprietary \ac{llm} developed by Anthropic.
The model is not open source, so its weights and architectural details cannot be inspected or deployed locally.
We access it through Anthropic's Claude Pro subscription plan.

\textbf{\ac{qwen}}~\cite{qwen} is an open-weight code-specialized \ac{llm} developed by Alibaba's Qwen team.
It is designed for agentic coding workflows, including program synthesis, debugging, code modification, and tool-based software development.
Unlike the proprietary models, it can be deployed locally without subscription or API fees, improving reproducibility and cost efficiency.
We use the 35B active-parameter version, which we run locally on three NVIDIA A100 40\,GB GPUs.

\textbf{OpenCode} is an open-source terminal-based coding agent that enables \ac{llm}-based software development directly within a local project environment.
It allows \acp{llm} to inspect files, modify source code, execute commands, and interact with external tools.
We use OpenCode \verb|1.14.20| and extend it with a Julia-documentation \ac{mcp} server~\cite{plavin_julia_mcp} that provides structured access to the official Julia documentation, enabling the agent to retrieve relevant API references and usage patterns during code generation.
In addition, we use a dedicated system prompt for Julia-based \ac{hpc} tasks.
The prompt defines the agent as an expert in Julia \ac{hpc} and emphasizes parallel programming, performance optimization, profiling, memory management, and efficient program design.
The agent is explicitly instructed to use \texttt{juliadoc} for documentation retrieval and to validate generated implementations.

Our \textbf{Sandbox} is an isolated Ubuntu 24.04 environment in which all code generation and execution are performed.
It is preconfigured with Julia, MPI, and Slurm to emulate realistic \ac{hpc} conditions, allowing the agent to autonomously generate, execute, and validate parallel and distributed workloads in a consistent software environment.
The isolated setup improves reproducibility and prevents side effects between experiments.

\subsection{Code Generation}\label{sec:setup_codegen}
To evaluate \ac{llm}-based agents on parallel task-based Julia code, we compare \texttt{Dagger.jl} with the established approaches \texttt{Base.Threads} and \texttt{MPI.jl}.
\texttt{Dagger.jl} is particularly relevant in this context because it supports task-graph-based execution and can be used in shared-memory, distributed-memory, or hybrid configurations.
For comparability, we distinguish between shared- and distributed-memory settings, as shared-memory \texttt{Dagger.jl} implementations may rely on different data structures and execution mechanisms than distributed ones.
The shared-memory \texttt{Dagger.jl} variant is compared against \texttt{Base.Threads}, while the distributed variant uses \texttt{Distributed.jl} and is compared against \texttt{MPI.jl}.
Hybrid implementations are excluded from the evaluation.

We analyze how effectively agents implement these algorithms across the programming models, comparing correctness, scalability, and performance.
A key design consideration is the trade-off between constraint enforcement and model flexibility: strict constraints improve reproducibility and reduce invalid implementations, but may limit solution strategies, while less restrictive prompts allow more independent decisions at the cost of correctness, efficiency, and comparability.
To balance both, our workflow uses two sequential stages in the same OpenCode session: test generation defines the interface and validation, and kernel generation implements the algorithm based on the generated test program.


In the first stage, a test program is generated which provides the execution harness for the actual kernel.
It defines how the kernel is invoked, generates and prepares the input data sequentially, collects the kernel output, and validates it using correctness checks. 
This setup separates test orchestration from the measured kernel execution while still leaving the concrete data-passing strategy between the test program and the kernel implementation.
Shared-memory implementations can usually operate on standard Julia arrays, whereas distributed-memory variants may require communication, data copies, or distributed abstractions such as \texttt{Dagger.jl}'s \texttt{DArray}.
Accordingly, the kernel interface is not fixed during test generation; instead, the \ac{llm} may define an interface suitable for the respective programming model and transform or distribute the input data before invoking the kernel.
For comparability, only the outer test-function interface is fixed: each test receives the number of experiments, the problem size, and the number of warmup runs, with an additional block-size parameter for task-based implementations to control decomposition granularity.
%

In the second stage, the agent implements the kernel in the same OpenCode session, using the generated test program as the interface, execution harness, and correctness checks.
Unlike the test-generation prompt, the kernel-generation prompt provides guidance for efficient, comparable implementations: external dependencies that directly solve the target problem are disallowed, while BLAS calls are permitted.
The agent is then instructed to test the implementation with the generated test program and kernel, run scaling experiments with appropriate problem sizes, and refine the code based on the results.
Correctness is assessed using the generated test program, while scaling experiments encourage performance-oriented improvements.

\section{Experimental Results}\label{sec:results}

To evaluate the ability of different \acp{llm} to generate parallel Julia code and to compare the established parallelization approaches with task-based \texttt{Dagger.jl}, we analyze both the generation process and the performance of the generated programs.
First, we examine the generation logs to identify differences between models, algorithmic problems, and Julia frameworks, including the time required for kernel generation.
Second, we execute the generated programs on the Otus supercomputer and evaluate their strong-scaling performance.
The dataset containing raw results, generation logs, and prompts is available at~\cite{repo_agentic_ai_julia_supercomputers}.
Each model--problem--framework combination is generated only once, and although generation is non-deterministic, the agentic workflow's iterative testing and refinement mitigates this variability.

The Otus supercomputer comprises 636 compute nodes, each with two AMD EPYC 9655 ``Turin'' CPUs, each with 96~cores at 2.6--4.5\,GHz, for a total of 192~cores and 768\,GiB of main memory per node.
Shared-memory experiments run on a single node with up to 192 cores, whereas distributed-memory experiments run on two nodes with up to 384 cores.
All experiments use Julia 1.12.6; \texttt{Dagger.jl} implementations use \texttt{Dagger.jl}~0.19.4, and MPI implementations use \texttt{MPI.jl}~0.20.26 with Open~MPI 5.0.7.
All runs are submitted as Slurm batch jobs and Slurm handles node allocation, process placement, and CPU binding.
In distributed runs, processes are spread evenly across nodes.
OpenBLAS is restricted to a single thread throughout all experiments to prevent nested parallelism from influencing the measured scaling behavior. 
Each experiment is repeated five times, and the reported results are averaged across these runs.

\subsection{Agentic AI for Julia}
Across the logs of \ac{llm}-generated Julia implementations, agents typically start with API exploration, then perform correctness testing, performance experiments, and iterative optimization.
Correctness is usually tested before evaluating performance across thread or process counts, problem sizes, and, where applicable, block sizes.
The logs mainly differ in structure: \ac{gpt} and \ac{opus} usually plan and proceed step by step, whereas \ac{qwen} more often uses direct trial-and-error and includes larger portions of generated or modified code.

The models also differ in their use of \texttt{Dagger.jl} abstractions.
\ac{gpt} often uses future-oriented constructs with deferred execution and explicit result resolution, while \ac{opus} more often expresses computations through explicit task dependencies or DAG-like execution patterns.
\ac{qwen} spans a wider range of abstraction levels, from high-level \texttt{Dagger.jl} interfaces down to lower-level or internal API interactions.
At the same time, even \ac{gpt} and \ac{opus} often prefer established implementation strategies over less familiar distributed data abstractions.
For example, when the prompt does not explicitly require \texttt{Dagger.jl}'s distributed data structures, \ac{gpt} may recognize \texttt{DArray} as suitable but still choose manual data transfers to worker processes.
Since such transfers can increase memory pressure in distributed-memory settings, we adapted the distributed \texttt{Dagger.jl} prompts to explicitly require its distributed data structures.
Some \ac{gpt} logs also attempt explicit task placement, using \texttt{Dagger.Scope} annotations or custom round-robin schemes.
The resulting error patterns reflect these abstraction choices: \ac{gpt} and \ac{opus} logs mainly show API adjustments, type corrections, data-distribution changes, and task-dependency fixes before reaching executable implementations.
In contrast, \ac{qwen} logs show more API errors, type errors, runtime crashes, and timeouts.
In several \ac{qwen} cases, implementations are adjusted by reducing the problem size or modifying the test configuration.

Performance-related decisions are mostly empirical across all \acp{llm}.
The logs contain limited evidence of an explicit model relating problem size, chunk size, worker count, scheduler overhead, and expected runtime.
Instead, the agents run benchmark experiments and adapt the implementation based on observed results.
\ac{opus} tends to test more extensively and often repeats correctness checks after performance experiments.
\ac{gpt} more frequently accepts limited local scaling results or notes that local results may not reflect larger-cluster behavior.
\ac{qwen} often uses comparatively small problem sizes despite prompt instructions to choose sizes with sufficiently long serial runtime, causing several measurements to be dominated by overhead.

\begin{figure}[t]
    \centering
    \begin{subfigure}{0.49\textwidth}
        \centering
        \includegraphics[width=\linewidth]{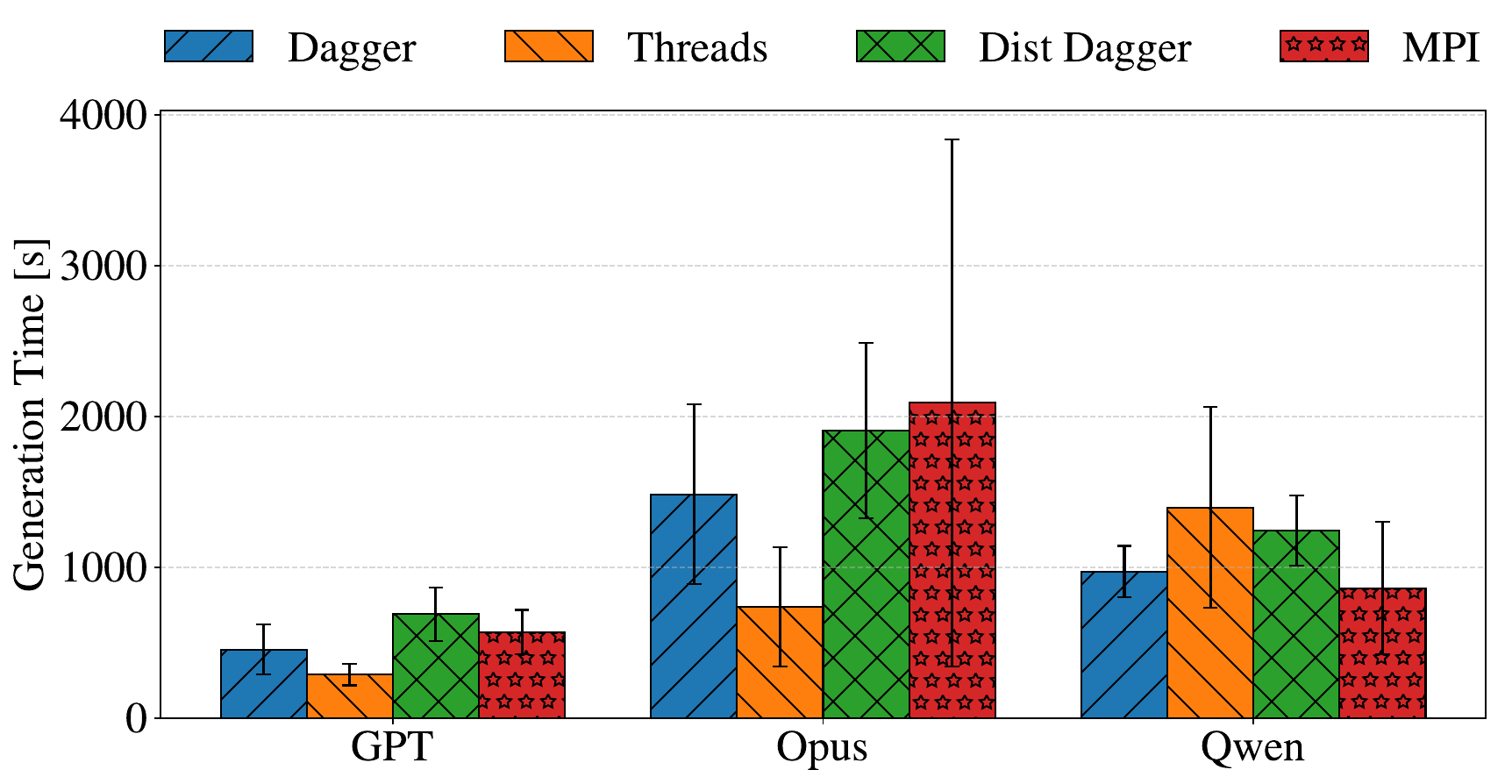}
        \caption{Averaged over different Problems}
        \label{fig:gen_framework}
    \end{subfigure}
    \hfill
    \begin{subfigure}{0.49\textwidth}
        \centering
        \includegraphics[width=\linewidth]{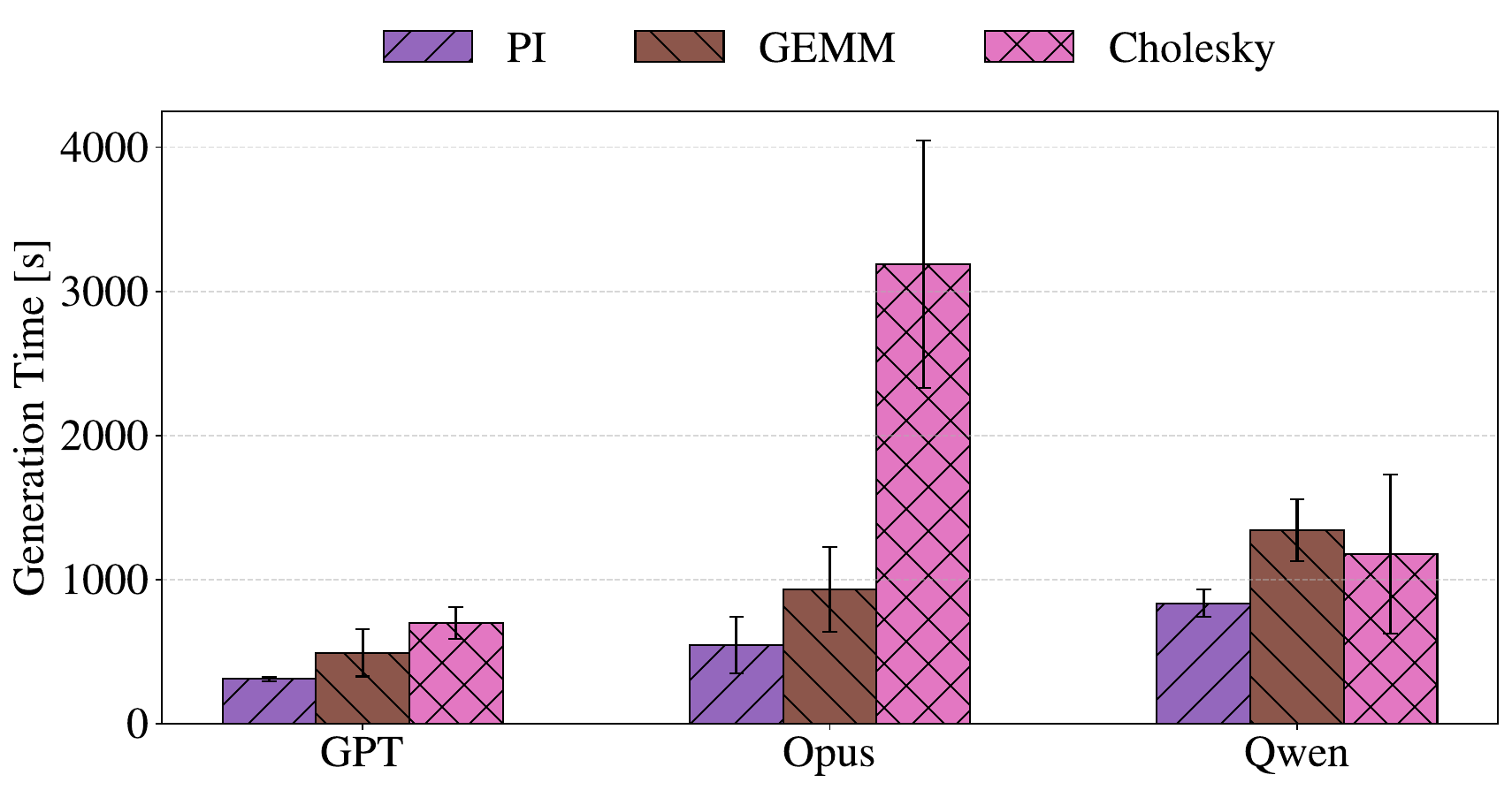}
        \caption{Averaged over Frameworks}
        \label{fig:gen_problem}
    \end{subfigure}
    \caption{Kernel generation time (code generation, testing, and optimization)}
    \label{fig:gen_time}
\end{figure}

\Cref{fig:gen_time} shows that kernel generation time varies substantially across algorithmic problems.
The $\pi$ kernel has the shortest generation times in most configurations, while the Cholesky kernel generally requires the longest, consistent with its more complex numerical dependencies and blocked execution structure.
\Cref{fig:gen_framework}, which averages over problems, shows lower generation times for \texttt{Base.Threads} than for the other frameworks, consistent with its comparatively direct fork-join structure.
Distributed-memory frameworks require additional decisions about data distribution, communication, task placement, and process coordination.
\Cref{fig:gen_problem}, averaging over frameworks, further shows that variation between algorithmic problems is larger than variation between frameworks.
Differences between \acp{llm} are also visible in the timing results.
\ac{opus} shows increased generation time for Cholesky, consistent with the more extensive testing and repeated validation observed in its logs, and \ac{qwen} shows a distinct timing profile compared to \ac{gpt} and \ac{opus}.
However, these timings must be interpreted alongside correctness results, since shorter generation times do not necessarily imply successful or correct implementations.
Because generation runs against hosted APIs and a local deployment, the measured times are affected by uncontrolled factors such as rate-limiting, token-throttling, and queueing.
Thus, we treat kernel generation time as an indicative proxy rather than a precise measurement.

\subsection{Performance Evaluation}
The generated test and kernel files enable agents to detect and correct common issues in \ac{llm}-generated code, including missing imports, interface mismatches, and simple runtime errors.
For small problem sizes and up to four threads or processes, all agents produced executable implementations for all algorithmic problems.
However, correctness differs between models: \ac{gpt} and \ac{opus} consistently generated implementations that passed the correctness checks for all problems, whereas \ac{qwen} implementations were executable in several cases but produced incorrect results in others.

In later experiments, some implementations failed beyond certain problem sizes or thread/process counts, with failure modes including deadlocks, oversubscription, out-of-memory errors, crashes, and timeouts.
This affected \ac{qwen} in particular: for problem sizes suitable for scaling experiments and successfully used with \ac{gpt} and \ac{opus}, its implementations failed for all problems, often already at small thread or process counts.
Therefore, the following performance evaluation plots include only \ac{gpt} and \ac{opus}.
Missing data points in the performance figures indicate configurations where the corresponding implementation failed at the specified number of threads or processes.

\begin{figure}[t]
    \centering
    \begin{subfigure}{0.49\textwidth}
        \centering
        \setlength{\abovecaptionskip}{-1pt}
        \includegraphics[width=\linewidth]{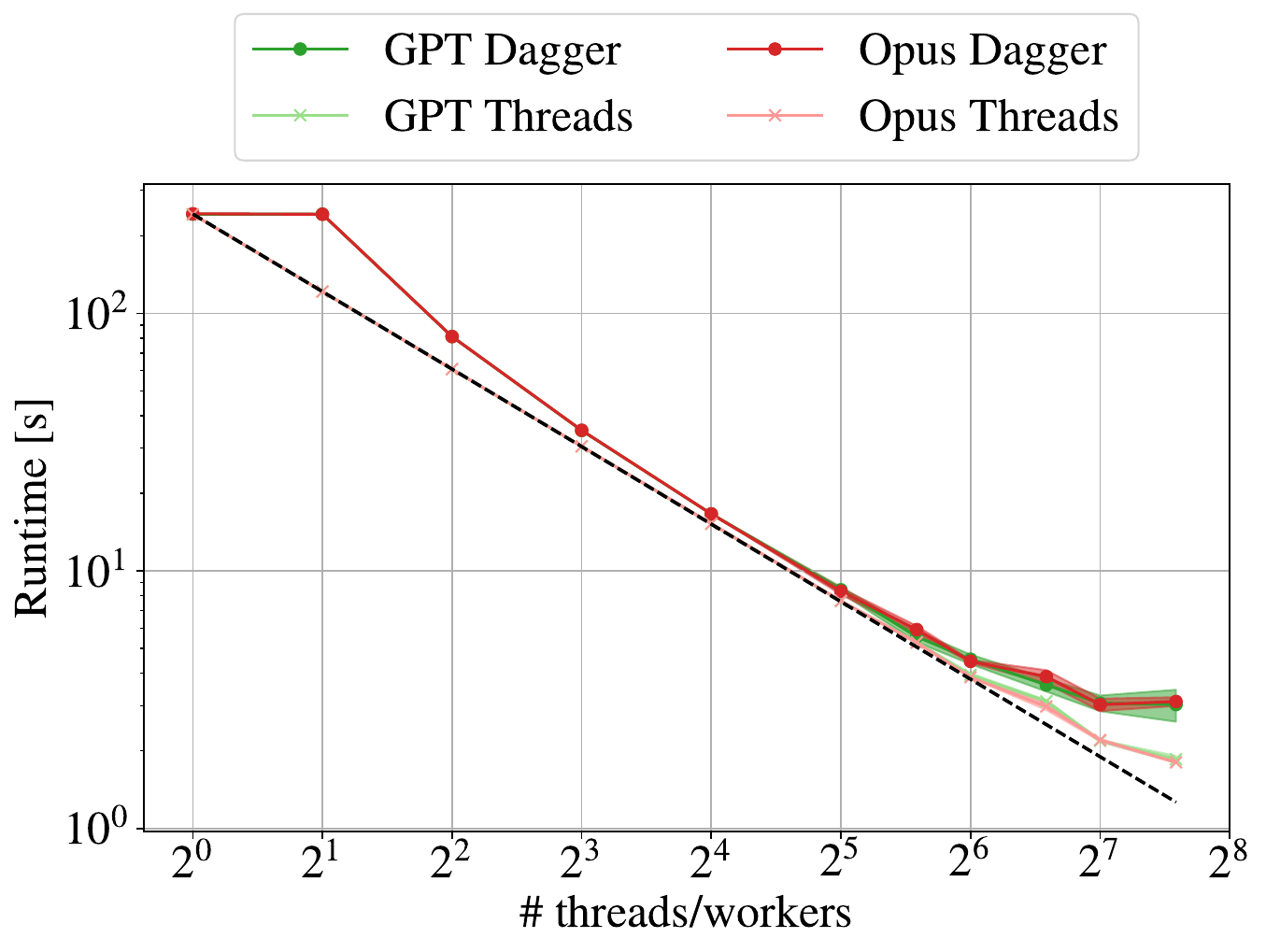}
        \caption{Runtimes in seconds on one node}
        \label{fig:shared_pi}
    \end{subfigure}
    \hfill
    \begin{subfigure}{0.49\textwidth}
        \centering
        \setlength{\abovecaptionskip}{-1pt}
        \includegraphics[width=\linewidth]{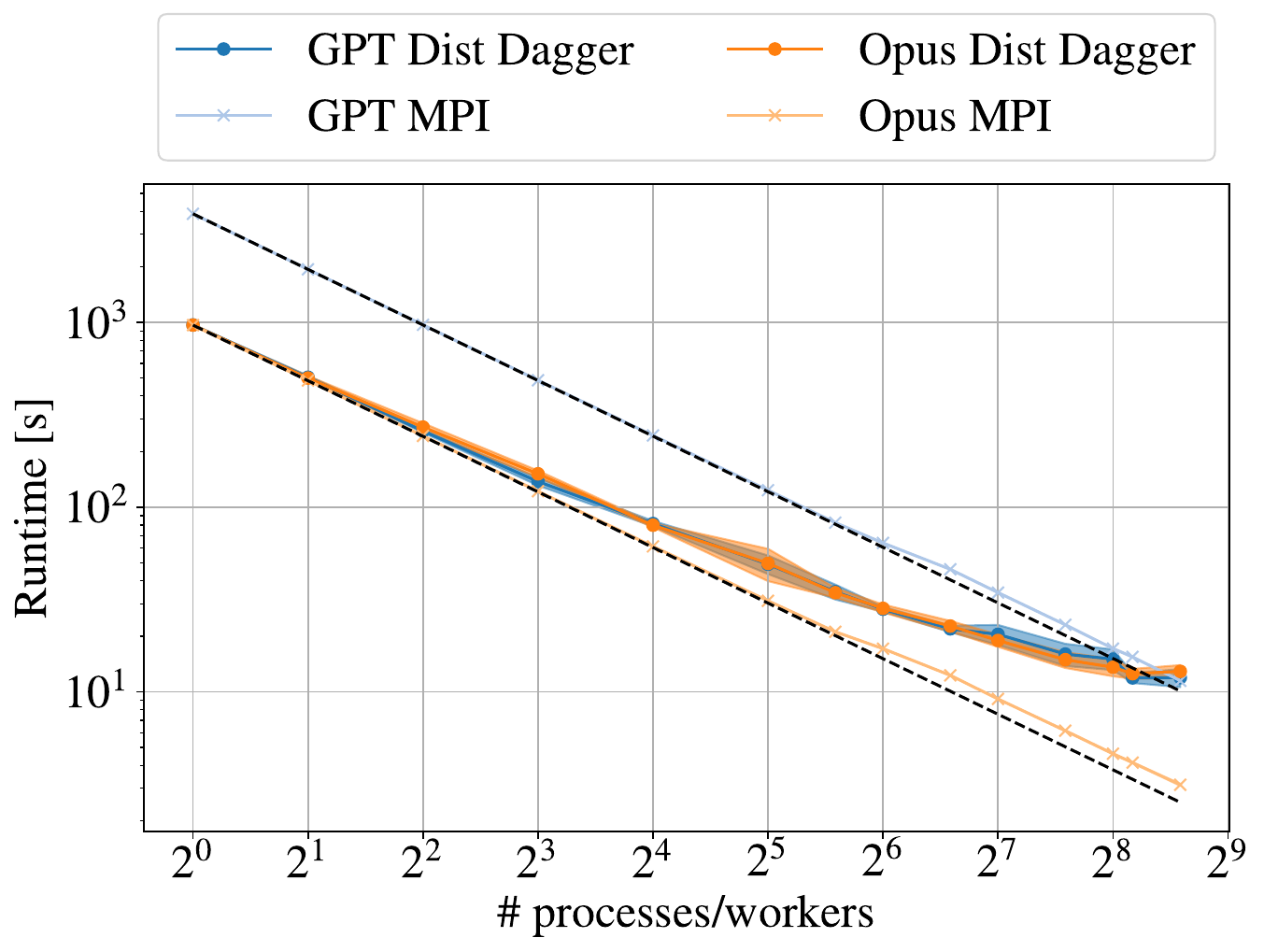}
        \caption{Runtimes in seconds on two nodes}
        \label{fig:dist_pi}
    \end{subfigure}
    \caption{Strong scaling of the $\pi$ approximation with $2^{41}$ intervals for shared- (left) and $2^{43}$ for distributed-memory (right).}
    \label{fig:pi}
\end{figure}

\textbf{Numerical $\pi$ Integration.}\label{sec:results_pi}
\Cref{fig:shared_pi} shows the shared-memory results.
\texttt{Base.Threads} is faster than \texttt{Dagger.jl} due to lower scheduling overhead.
Runtime decreases with more threads, but scaling efficiency drops at higher parallelism.
Both \acp{llm} achieve comparable performance and near-ideal scaling.

A noticeable discontinuity occurs in the \texttt{Dagger.jl} runtime between one and two threads.
This is caused by \texttt{Dagger.jl} reserving one \emph{interactive} thread for task scheduling, leaving only $\text{num\_threads} - 1$ threads for worker execution.
As a result, the effective worker count is shifted relative to the \texttt{Base.Threads} implementation.
This behavior applies to all shared-memory \texttt{Dagger.jl} implementations discussed in the following.
As explained in \Cref{sec:setup_codegen}, the generated \texttt{Dagger.jl} implementations differ between shared- and distributed-memory settings, so their runtimes may differ as well.

\Cref{fig:dist_pi} presents the distributed-memory results.
For this embarrassingly parallel workload, MPI achieves near-ideal scaling, since only a final reduction is required.
Distributed \texttt{Dagger.jl} also scales well initially, but task-management overhead reduces efficiency at higher process counts.
Both \acp{llm} generate reduction strategies that avoid naive centralized communication and therefore achieve good distributed scaling behavior.
\ac{qwen} generated a distributed \texttt{Dagger.jl} implementation based on a \verb|DArray| decomposition, which introduces substantial communication and allocation overhead.
Therefore, its runtime is orders of magnitude higher than that of \ac{gpt} and \ac{opus}.
The runtime difference between the \ac{gpt} and \ac{opus} MPI implementations is caused by differences in the generated kernel code.
\ac{opus} relies on a single SIMD-annotated loop, whereas \ac{gpt} uses manual loop unrolling, which performs worse in this case.






\textbf{General Matrix Multiplication.}\label{sec:results_gemm}
A tiled GEMM kernel is, similar to the $\pi$ benchmark, a highly structured workload with regular communication patterns.
At the same time, it is in principle well suited to task-based execution, since the multiplication can be decomposed into tile-level tasks with explicit data dependencies and substantial per-task computation.
With sufficiently large tiles, this structure provides enough arithmetic intensity to amortize scheduling overhead.

\begin{figure}[t]
    \centering
    \begin{subfigure}{0.49\textwidth}
        \centering
        \setlength{\abovecaptionskip}{-1pt}
        \includegraphics[width=\linewidth]{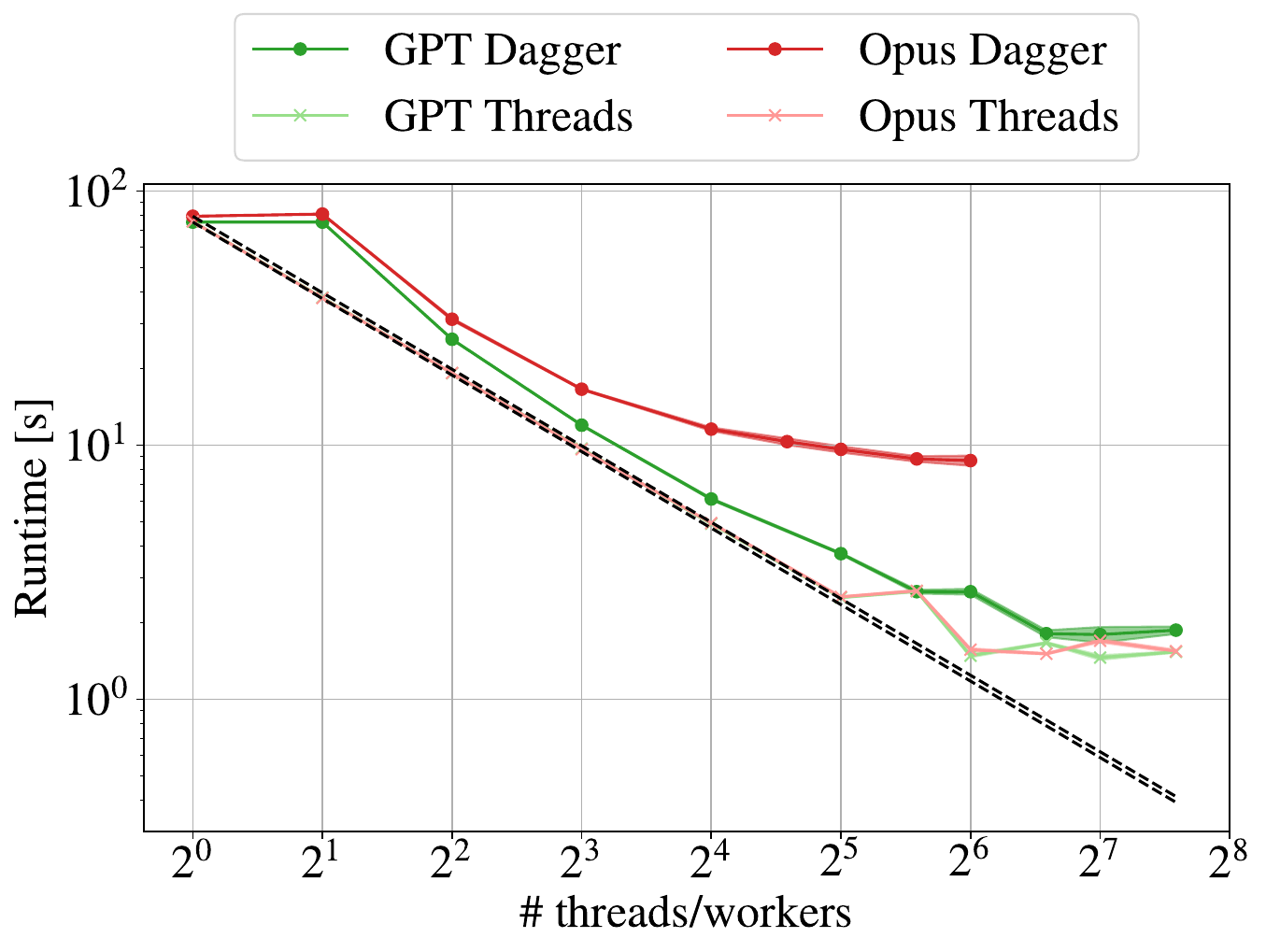}
        \caption{Runtimes in seconds on one node}
        \label{fig:shared_gemm}
    \end{subfigure}
    \hfill
    \begin{subfigure}{0.49\textwidth}
        \centering
        \setlength{\abovecaptionskip}{-1pt}
        \includegraphics[width=\linewidth]{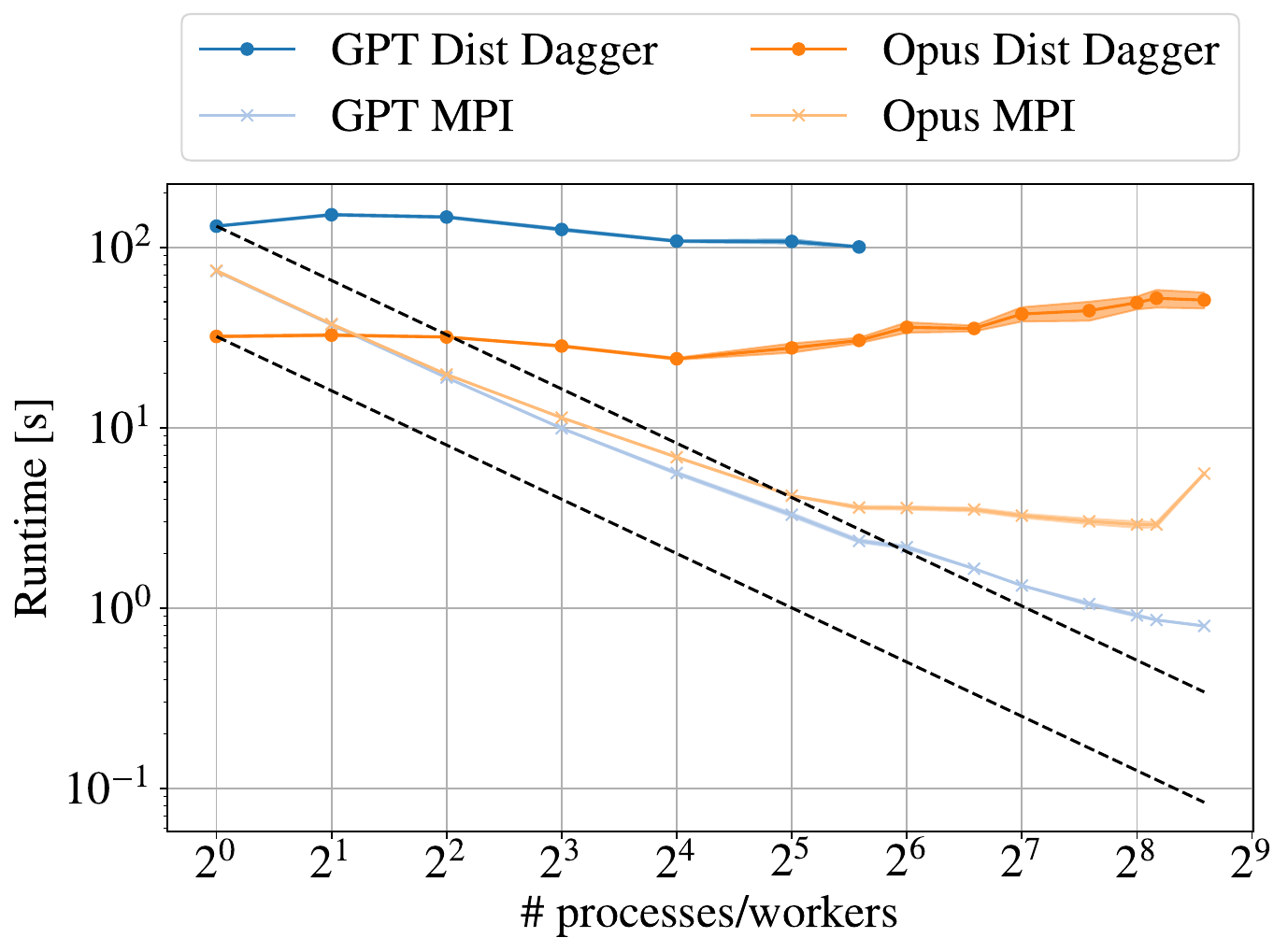}
        \caption{Runtimes in seconds on two nodes}
        \label{fig:dist_gemm}
    \end{subfigure}
    \caption{Strong scaling of the tiled GEMM for $2^{14}\times2^{14}$ matrices with tile size $2048$ in shared- (left) and distributed-memory (right) settings.}
    \label{fig:gemm}
\end{figure}

As shown in \Cref{fig:shared_gemm}, all implementations initially exhibit comparable performance.
However, as thread count increases, the \ac{opus} \texttt{Dagger.jl} implementation degrades more rapidly and fails to complete beyond 32 threads, indicating an unresolved scheduling dependency.
This suggests that the generated synchronization and dependency structure does not scale reliably to larger task graphs, despite the algorithmic structure's suitability for task-based parallelism.
Inspection of the generated implementations indicates that this behavior is largely related to scheduling overhead and dependency management.
The \ac{opus} \texttt{Dagger.jl} implementation relies on \texttt{spawn\_datadeps()}, whereas \ac{gpt} uses a simpler future-based approach in which asynchronously spawned tile computations are stored as task handles and synchronized explicitly after task creation.
Since tiled GEMM requires coordination both before computation, to distribute matrix tiles, and after computation, to collect partial results, the additional dependency tracking introduced by \texttt{spawn\_datadeps()} increases overhead and appears to limit scalability in the generated implementation.

The distributed-memory results in \Cref{fig:dist_gemm} show an even clearer separation between programming models.
The MPI-based implementations scale reasonably well with increasing process counts, whereas the distributed \texttt{Dagger.jl} implementations exhibit little to no scalability.
Both \ac{gpt} and \ac{opus} again make extensive use of \texttt{spawn\_datadeps()}.
For this regular workload, where communication and synchronization patterns can be expressed relatively directly, the generated dependency-based \texttt{Dagger.jl} implementations introduce more coordination overhead than the MPI implementations, even though the latter use a naive column-wise output decomposition rather than established, optimized GEMM algorithms.

\begin{figure}[t]
    \centering
    \begin{subfigure}{0.49\textwidth}
        \centering
        \setlength{\abovecaptionskip}{-1pt}
        \includegraphics[width=\linewidth]{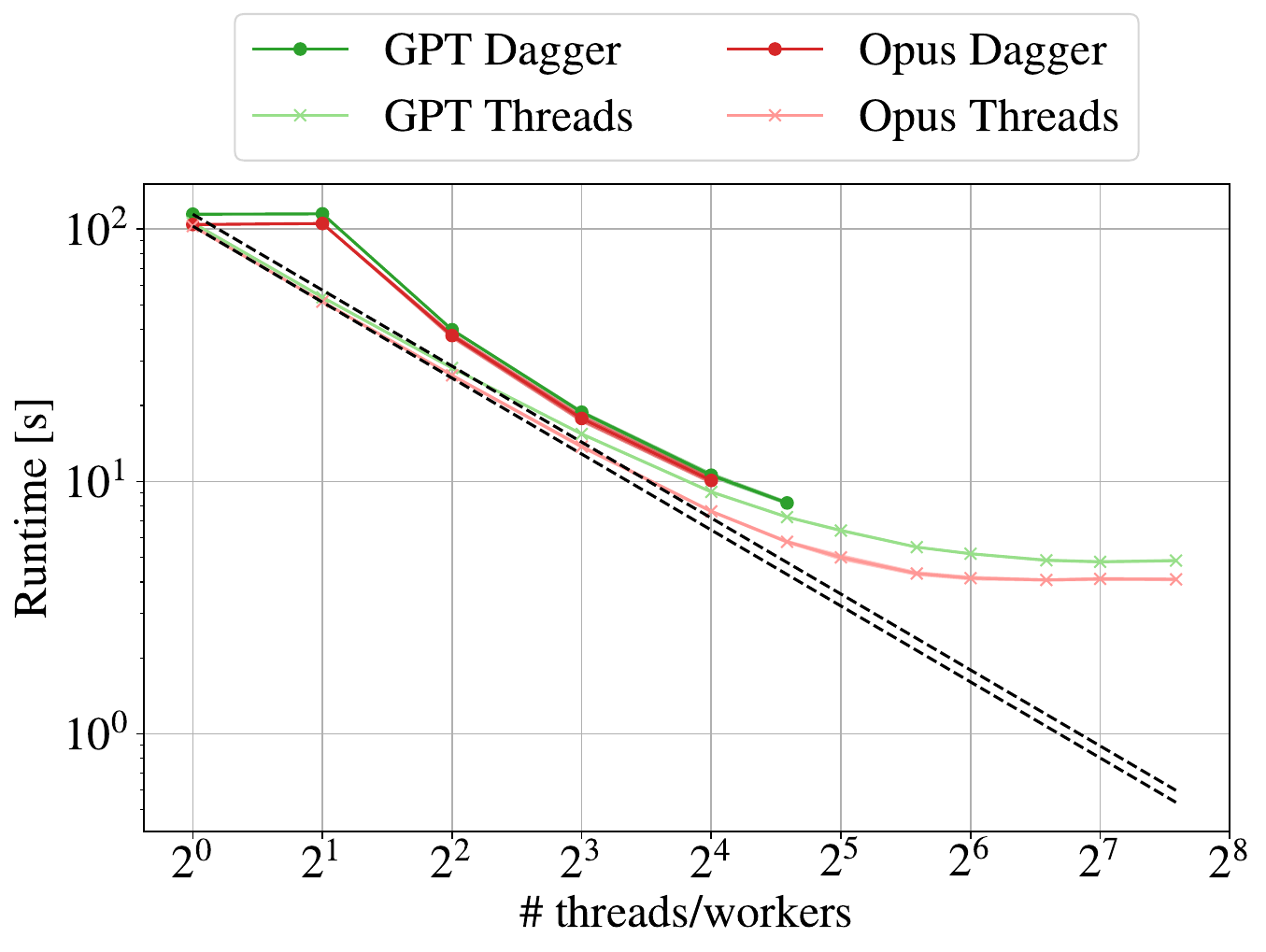}
        \caption{Runtimes in seconds on one node}
        \label{fig:shared_chol}
    \end{subfigure}
    \hfill
    \begin{subfigure}{0.49\textwidth}
        \centering
        \setlength{\abovecaptionskip}{-1pt}
        \includegraphics[width=\linewidth]{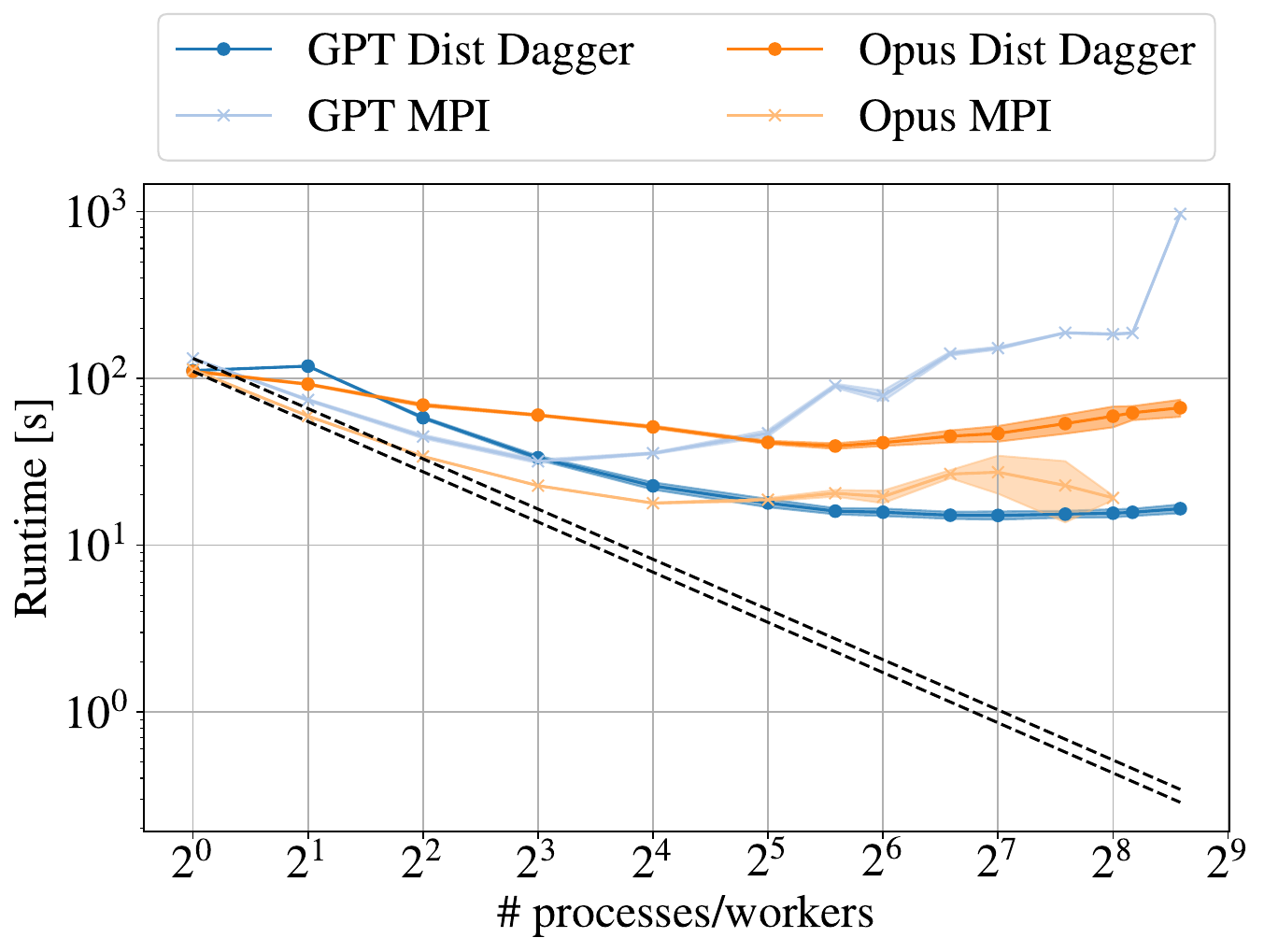}
        \caption{Runtimes in seconds on two nodes}
        \label{fig:dist_chol}
    \end{subfigure}
    \caption{Strong scaling of the tiled Cholesky decomposition for a $2^{14}\times2^{14}$ matrix with tile size $2048$ in shared- (left) and distributed-memory (right) settings.}
    \label{fig:chol}
\end{figure}

\textbf{Cholesky Decomposition.}\label{sec:results_chol}
The tiled Cholesky decomposition has an irregular dependency structure and is therefore a common benchmark for task-based runtime systems.
For the shared-memory implementations, \Cref{fig:shared_chol} shows behavior similar to the \ac{opus} GEMM implementation: the \ac{gpt} and \ac{opus} \texttt{Dagger.jl} variants encounter deadlocks at 24 and 32 tasks, respectively.
In contrast, the fork-join-based \texttt{Base.Threads} implementations run up to 192 threads, with scaling saturating around 64 threads.

Inspection of the shared-memory \texttt{Dagger.jl} implementations shows that both models express the algorithm through futures but organize dependency tracking differently.
\ac{opus} wraps the decomposition in a single \texttt{spawn\_datadeps()} region and encodes dependencies explicitly using \texttt{In} and \texttt{InOut} annotations, relying on a globally managed dependency scope.
\ac{gpt} instead stores the futures returned by \texttt{@spawn} directly in the chunk array and uses them to construct dependencies implicitly, resulting in a more local scheduling structure without an additional global dependency region.

The same distinction appears in the distributed-memory variants: \ac{opus} again uses \texttt{spawn\_datadeps()} with annotated accesses, whereas \ac{gpt} relies on future-based composition.
The distributed MPI and \texttt{Dagger.jl} implementations show comparable performance at low process or worker counts, with only limited overall scaling.
At higher parallelism levels, the \texttt{Dagger.jl} implementations exhibit smoother scaling, whereas MPI shows more irregular behavior.

\section{Discussion}\label{sec:discussion}
Our experimental results in \Cref{sec:results} show that agentic code generation can detect and correct many common issues in \ac{llm}-generated code, such as missing imports, interface mismatches, and simple runtime errors, since programs are executed and tested by the agent during generation.
As a result, agents generally produce executable parallel Julia code for small problem sizes and low degrees of parallelism.
However, in an \ac{hpc} context, correctness is only a necessary condition; robustness, scalability, and performance on supercomputers matter equally.
Several implementations that pass small correctness tests fail when problem size or parallelism is increased, for example, due to deadlocks, oversubscription, timeouts, or out-of-memory errors.
This shows that the generated parallel code must be evaluated under scaling conditions, not only through functional tests.

The model comparison further shows that generation logs are essential for interpreting the results.
\ac{gpt} and \ac{opus} follow a more structured development process involving planning, correctness testing, and performance experiments, with \ac{opus} testing particularly extensively, whereas \ac{qwen} shows less structured behavior and more often uses problem sizes that are too small for meaningful scaling experiments.
Overall, the implementations generated by \ac{gpt} and \ac{opus} perform similarly, but \ac{gpt} achieves better performance and scalability in most \texttt{Dagger.jl} cases.
A possible explanation is that \ac{opus} spends more effort optimizing and testing code on the local execution system, which may improve performance in the observed test environment but can lead to implementations that are overly tuned to it and therefore transfer less effectively to supercomputers.
This points to a broader limitation: during generation, agents lack access to distributed-memory execution, realistic process placement, node-level memory constraints, and large core counts, making it difficult to predict performance on \ac{hpc} systems, especially for distributed and task-based implementations in which data distribution, communication, task granularity, synchronization, and scheduling overhead are decisive.
Ideally, agents would optimize directly on the target system, where these effects become observable.
In practice, however, this is difficult because supercomputers are accessed through batch schedulers, offer limited interactive feedback, and repeated agent-driven benchmark runs would consume substantial shared resources.
If this barrier were lowered, for example through agent-accessible batch submission, agents could optimize directly on the target system, likely narrowing the observed gap between locally tuned and supercomputer behavior, especially for distributed and task-based implementations.
The trade-off then becomes economic, weighing the cost of repeated on-system runs against the resulting gains.

The results also reveal a tension between task-based programming models and the agents' implementation strategies.
Although frameworks such as \texttt{Dagger.jl} delegate scheduling and data movement to the runtime, generated implementations often reintroduce manual control through explicit task placement, custom scheduling, or manual data transfer.
Agents also favor familiar data structures and execution patterns over framework-specific abstractions.
Thus, the results do not show that task-based execution is unsuitable for these workloads, but that agents do not always exploit the runtime effectively.

The prompt adaptations used in this study demonstrate that explicit constraints are necessary for fair comparisons across frameworks.
If prompts are too permissive, agents may use a framework syntactically correctly without adhering to its intended programming model, for example by manually moving data in distributed \texttt{Dagger.jl} implementations instead of using distributed data abstractions.
If prompts are too restrictive, they may suppress framework-specific solution strategies.
Prompt design therefore requires a trade-off between comparability, reproducibility, and implementation flexibility.

The models differ in cost, though our data supports only a qualitative assessment: the proprietary \ac{gpt} and \ac{opus} incur subscription fees, whereas the open-weight \ac{qwen} runs locally without per-token cost but requires dedicated GPU hardware.
Kernel generation time (\Cref{fig:gen_time}) is the only cost dimension we measure directly.
A full cost--time--value analysis would additionally require token counts and human guidance time, which we leave to future work.

Finally, two limitations stand out.
Our comparison lacks a non-Julia ground truth: \texttt{Base.Threads} and \texttt{MPI.jl} act as within-Julia baselines, but a hand-written or vendor-optimized reference would be needed to separate framework-inherent inefficiencies from those of the generated code.
We also do not trace implementation choices to training sources.
The full logs and prompts are available~\cite{repo_agentic_ai_julia_supercomputers}.

\section{Conclusion}\label{sec:conclusion}
This work evaluated \ac{llm}-based agents for generating parallel Julia code with \texttt{Base.Threads}, \texttt{MPI.jl}, and \texttt{Dagger.jl}.
The results show that Agentic AI can produce executable implementations and automatically correct many typical coding errors through iterative testing.
However, small-scale correctness does not guarantee robustness or scalability in an \ac{hpc} setting.
Several implementations fail at larger problem sizes or higher thread and process counts, exposing deadlocks, oversubscription, or out-of-memory errors.
This highlights the need to evaluate and optimize generated \ac{hpc} implementations under realistic conditions.

Overall, the agents handle established models such as \texttt{Base.Threads} and \texttt{MPI.jl} more reliably than task-based \texttt{Dagger.jl} implementations, which require more complex decisions about task granularity, data distribution, dependencies, and scheduling.
Thus, \ac{llm}-based agents are promising for producing initial parallel code, but generating robust, performance-aware implementations for large-scale \ac{hpc} systems remains an open challenge.
Future work should investigate how agentic optimization can be coupled more tightly to the target system, for example through batch-aware feedback loops and resource-conscious benchmarking.

\ifanonymous
\else
    \begin{credits}
        \subsubsection*{\ackname}
        \ifanonymous
        Acknowledgments omitted for double-blind review.
        \else
        This research was supported by Advantest as part of the Graduate School ``Intelligent Methods for Test and Reliability'' (GS-IMTR) at University of Stuttgart.
        This research was partially funded by the Deutsche Forschungsgemeinschaft (DFG, German Research Foundation) under \href{https://gepris.dfg.de/gepris/projekt/558599020?language=en}{project number~558599020}.

        The authors gratefully acknowledge the computing time made available to them on the high-performance computer \href{https://pc2.uni-paderborn.de/systems-and-services/otus}{Otus at the NHR Center Paderborn Center for Parallel Computing (PC2)}.
        This center is jointly supported by the Federal Ministry of Research, Technology and Space and the state governments participating in the National High-Performance Computing (NHR) joint funding program.
        \fi

        \subsubsection*{\discintname}
        The authors have no competing interests to declare that are relevant to the content of this article.
    \end{credits}
\fi
\bibliographystyle{splncs03_unsrt_etal}
\bibliography{references}

\end{document}